\documentclass[twocolumn,twocolappendix]{aastex63} %

\newcommand{\code}[1]{\texttt{#1}}
\newcommand{\rft}{\code{REFITT}}
\usepackage{afterpage}
\usepackage{hyperref}
\usepackage{amsmath}

\shorttitle{\rft}
\shortauthors{Sravan et al.}

\begin{document}


\title{Real-Time Value-Driven Data Augmentation in the Era of LSST}


\author{Niharika Sravan}
\affiliation{Department of Physics and Astronomy\\Purdue University\\525 Northwestern Ave., West Lafayette, IN 47907, USA}
\author{Dan Milisavljevic}
\affiliation{Department of Physics and Astronomy\\Purdue University\\525 Northwestern Ave., West Lafayette, IN 47907, USA}
\author{Jack M. Reynolds}
\affiliation{Department of Physics and Astronomy\\Purdue University\\525 Northwestern Ave., West Lafayette, IN 47907, USA}
\author{Geoffrey Lentner}
\affiliation{Department of Physics and Astronomy\\Purdue University\\525 Northwestern Ave., West Lafayette, IN 47907, USA}
\author{Mark Linvill}
\affiliation{Department of Physics and Astronomy\\Purdue University\\525 Northwestern Ave., West Lafayette, IN 47907, USA}

\submitjournal{The Astrophysical Journal}



\begin{abstract}
The deluge of data from time-domain surveys is rendering traditional human-guided data collection and inference techniques impractical.
We propose a novel approach for conducting data collection for science inference in the era of massive large-scale surveys that uses value-based metrics to autonomously strategize and co-ordinate follow-up in real-time.
We demonstrate the underlying principles in the Recommender Engine For Intelligent Transient Tracking (\rft) that ingests live alerts from surveys and value-added inputs from data brokers to predict the future behavior of transients and design optimal data augmentation strategies given a set of scientific objectives.
The prototype presented in this paper is tested to work given simulated Rubin Observatory Legacy Survey of Space and Time (LSST) core-collapse supernova (CC SN) light-curves from the PLAsTiCC dataset.
CC SNe were selected for the initial development phase as they are known to be difficult to classify, with the expectation that any learning techniques for them should be at least as effective for other transients.
We demonstrate the behavior of \rft\ on a random LSST night given $\sim$32000 live CC SNe of interest. 
The system makes good predictions for the photometric behavior of the events and uses them to plan follow-up using a simple data-driven metric.
We argue that machine-directed follow-up maximizes the scientific potential of surveys and follow-up resources by reducing downtime and bias in data collection. 
\end{abstract}


\section{Introduction}

A new paradigm of astronomical data collection has been realized over the past decade.
Dozens of projects\footnote{e.g. ASSA-SN; \citet{2017PASP..129j4502K}, ATLAS; \citet{2018PASP..130f4505T}, CRTS; \citet{2011arXiv1102.5004D}, DLT40; \citet{2017ApJ...848L..24V}, KAIT; \citet{2001ASPC..246..121F}, 
PanSTARRS; \citet{2016arXiv161205560C}, ZTF; \citet{2014htu..conf...27B, 2017NatAs...1E..71B}} probing the time-variable universe have led to new discoveries \citep[e.g.,][]{2016ApJ...819...35A,2017ApJ...836...60L,2019ApJ...872...18M,2019MNRAS.484.1031P}, higher quality data-sets and statistics \citep[e.g.,][]{2017ApJ...837..121G, 2019ApJ...874..106B} and improvements in our understanding of the physics governing transient phenomena.
The advent of the Rubin Observatory Legacy Survey of Space and Time (LSST) \citep{2002SPIE.4836...10T,2008arXiv0805.2366I,2019ApJ...873..111I} in the upcoming decade, delivering an order of magnitude higher volume of alerts, will be transformative.
However, several limitations prevent us from fully exploiting the science potential of even current survey output. 

A major obstacle is the real-time filtration, interpretation, and prioritization of survey alert streams to identify candidates for follow-up. 
Several data brokers (e.g., ALeRCE\footnote{\url{http://alerce.science/}}; AMPEL [\citealt{2019A&A...631A.147N}]; Antares [\citealt{2014SPIE.9149E..08S}]; Lasair [\citealt{2019RNAAS...3a..26S}]; MARS\footnote{\url{https://mars.lco.global/}}; Pitt-Google Broker\footnote{\url{https://github.com/mwvgroup/Pitt-Google-Broker}}) 
have emerged recently with the goal of helping sort, cross-reference, and value-add survey alert streams \citep[e.g.,][]{2018ApJS..236....9N}, with each broker having different scientific priorities and scope of what they will provide to their subscribers.
Broker services are also supported by recent development of software infrastructure, known as Target and Observation Managers (TOMs) or marshals \citep[e.g.][]{2018SPIE10707E..11S,2019PASP..131c8003K}, that permit observers to sort through raw and broker alert streams to plan and trigger follow-up.
Although they represent important components of the overall cyberinfrastructure required to support survey science \citep[e.g.][]{2019NatRP...1..600H}, these services alone will be unable to truly maximize the science potential of the full LSST survey alert stream as they still largely rely on humans to make decisions. 

A second limitation is availability of follow-up resources.
Full characterization of transients often requires spectroscopy and/or mutli-wavelength and multi-messenger follow-up, but the availability of facilities remains relatively low, time-sensitive, and time-consuming.
ZTF's Bright Transient Survey partly addresses this issue by using a dedicated low-resolution IFUS 
to obtain classification spectra for most ZTF transients $<$18.5 mag \citep{2019PASP..131g8001G,2019arXiv191012973F}.
However, LSST will demand efficient value-based follow-up decisions that are adaptable to the diverse needs of potentially competing science goals (e.g., obtaining flash spectroscopy or planning UV/radio follow-up) for a much more massive alert stream.

The third limitation is that LSST's light-curve (LC) sampling will be insufficient to {\it a priori} be interpreted with theoretical models for many transients.
LSST will conduct two surveys, Wide Fast Deep (WFD) and Deep Drilling Field (DDF), in six filters (bands): u, g, r, i, z, and y.
However, $\sim 90$\% of LSST's time will be spent on the WFD survey, in which LSST is nominally expected to revisit the same field, in any band in $\sim$3 days and in the same band in $\sim$3 weeks \citep{2017arXiv170804058L}. 
Algorithms developed in response to the Photometric LSST Astronomical Time-Series Classification Challenge \citep[PLAsTiCC,][]{2018arXiv181000001T, 2019PASP..131i4501K} have made significant advances in developing classification models for LSST LCs \cite[e.g.,][]{2019AJ....158..257B,2019arXiv190905032G}.
Early-time classification using partial LCs is also possible \citep{2018arXiv180703869C,2020MNRAS.491.4277M,2019PASP..131k8002M}.
However, it is unclear to what extent LSST's LCs can be used to derive other types of science (e.g. explosion physics, stellar progenitor information).

We propose augmenting LSST's coverage with supporting facilities in real-time using a value-based metric for constraining science of interest.
The challenge is to shift the burden of follow-up decisions from observers to machines that are both scalable and not subject to human error, fatigue, or bias.
As a step toward this goal, in this paper we introduce the Recommender Engine For Intelligent Transient Tracking (\rft), an autonomous system that ingests live alerts from surveys and value-added inputs from brokers, predicts their future behavior, and makes recommendations for the most valuable events and epochs for obtaining additional photometry, given the goal of most optimally augmenting the alert stream for science inference.

This paper is organized as follows. 
In Section \ref{s:sol} we provide an overview of the key elements needed for conducting value-driven data augmentation.
In Section \ref{s:refitt} we present our implementation of these elements in \rft.
In Section \ref{s:results} we simulate the performance of \rft\ for LSST, illustrating its predictions and follow-up recommendations.
We discuss steps to improve \rft\ and conclude in Section \ref{s:conclusions}.

\section{Machine-Driven Data Augmentation} \label{s:sol}

We designate the class of autonomous systems that strategize and coordinate value-driven follow-up in real-time to be ORACLEs (Object Recommender for Augmentation and Coordinating Liaison Engine).
In this section, we identify three essential components of any ORACLE. 

\begin{figure*}
\begin{center}
\includegraphics[width=0.75\textwidth]{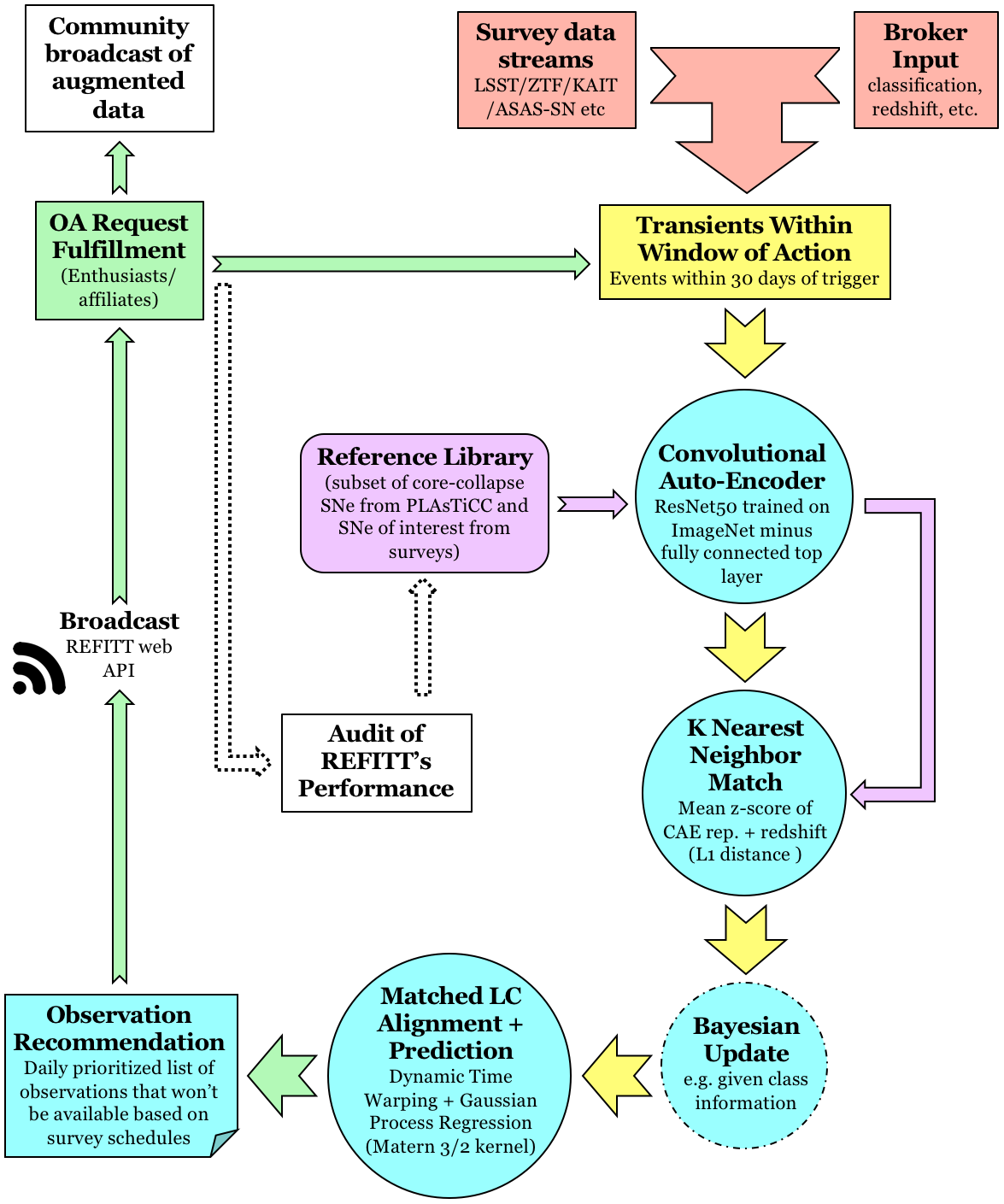}
\end{center}
\vspace{-10pt}
\caption{
Overview of the \rft\ system. 
It ingests data streams from surveys and inputs from data brokers and designs a follow-up strategy for optimally augmenting transient LCs for science inference.
Blue circles indicate \rft's AI components (discussed in \ref{s:ai}). 
Size of arrows indicate volume of data flow while their colors (red to green) represent their information entropy (high to low). OAs (observing agents) collect data recommended by the system which are then fed back into the alert stream.
OA submissions that deviate significantly from the system's expectations are assimilated into libraries it uses to make predictions.
\label{f:architecture}}
\end{figure*}

\begin{enumerate}
    \item {\it Signal prediction}: Decisions about the utility of a measurement require value estimates. Given a time-series signal, future values can be estimated using theoretical or regression models. Regression models have the dual advantage of being flexible and not requiring an understanding of the signal generating processes. Machine learning regression models capitalize on past data for this purpose and have been demonstrated to be very successful at a wide variety of forecasting problems \citep[e.g.,][]{WANG2015980,2018SoPh..293...28F,2019CQGra..36h5005M}.
    \item {\it Utility estimation}: Next is translation of signal expectation to the utility of obtaining a measurement. While simple and intuitive utilities can be chosen (e.g. utility proportional to prediction error to favor measurements at epochs with most uncertainty), in some cases it might be more appropriate to define measurement utility in relation to a target scientific model. Optimal design of experiments offers several utility functions for the goal of optimally constraining scientific models with respect to various statistical criteria  (e.g. A- or D-optimality). They can be used to design follow-up (experiments) that most optimally constrains theoretical models of interest \citep[see e.g.,][]{2015arXiv150102467Y}.
    \item {\it Data acquisition}: Finally, data must be collected in order of high expected utility up to a collection budget.
    While a fixed cost of measurement can be assumed, more sophisticated implementations require consideration of variable measurement cost (e.g. time for acquisition). For the former, given a set of observing agents (OAs), data collection tasks can either be allocated randomly or on first-come first-served basis. However, given strong differences between OA abilities to acquire data allocated, even elementary matching to drive task allocation (e.g. considering weather at an OA site) would lead to a significant improvement in the quality of data obtained.
\end{enumerate}

\rft\ is a prototype of an ORACLE.
It is a demonstration of the implementation of the above elements and seeks to provide a complete solution to the challenges of conducting real-time value-driven augmentation.

\newpage
\section{Recommender Engine for Intelligent Transient Tracking} \label{s:refitt}
\subsection{System Overview} \label{ss:sys}

Figure \ref{f:architecture} shows a schematic of the \rft\ system.
It ingests data streams from surveys and brokers and narrows down ongoing events within a chosen window of action.
In this paper, the window of action is set at 30 days since LSST trigger.
CC SN events (SNe II and SNe Ibc) from the PLAsTiCC database have been used to simulate the LSST alert stream.
CC SN types are known to be difficult to characterize due to their fast temporal evolution and, for some subtypes, similarity to SNe Ia. 
The choice keeps the scope of the initial development focused but ensures that solutions can be generalized to other transient types.

Candidates within the window of action are passed to \rft's AI for signal prediction. 
At its core, \rft\ relies on a convolutional auto-encoder (CAE) to match live events against a library of simulated events. For every live event, `k' most similar events in the reference library are chosen to predict future photometric behaviour. 
If available, classification information from brokers is used to update the system's choice for similar library events.
The `k' reference events are then used to estimate future photometric evolution of each event using a combination of Dynamic Time Warping alignment and Gaussian Process regression. 
We discuss each of the above components in detail in Section \ref{s:ai}. 

Next, all events within the window of action are assigned a quality score based on prediction confidence. 
In this paper, we use a simple data-driven metric to determine if follow-up is needed at the next observing epoch and prioritize observations by prediction quality scores.
The follow-up recommendation strategy is discussed in detail in Section \ref{s:recommendation}.

Finally, a web API distributes a list of targets to OAs supporting \rft\ according to expected photometric behavior of the target and individual OA's suitability for obtaining the observation (location, weather, aperture, instrumentation, etc).
Observations returned by OAs are fed back into the alert stream.
Events whose behavior differs significantly from the system's expectations are audited and assimilated into its reference library. 
We leave an exploration of strategies for optimal training set assembly for future work \citep[but see,][]{2019MNRAS.483....2I}.

\subsection{\rft\ AI} \label{s:ai}

\subsubsection{Reference library} \label{ss:lib}
\rft\ uses a library of events with full photometric information within the window of action (30 days since LSST trigger for this paper) to predict the behavior of live events. 
The library is built from a subset of CC SNe from the full PLAsTiCC dataset (training and testing) that satisfy each of the following criteria: 
\begin{enumerate}
    \item in a given LSST band there exists at least one photometric measurement before and after peak flux with mean flux value between a third to two-thirds of the mean peak flux value,
    \item the above is true in at least three bands, and 
    \item the above are satisfied within a 100 day window, 20 before and 80 days after LSST trigger.
\end{enumerate}
An LSST trigger is defined as the first photometric measurement in a time series where flux/$\sigma_{\rm flux} >$ 5 \citep[][]{2019PASP..131i4501K}.
Photometry with $\sigma_{\rm flux} >$ 50 units is removed.
These criteria ensure that conducting Gaussian Process regression on events in the reference library results in clean fits throughout the window of action. 
Performing the above selection on the PLAsTiCC dataset yields 3228 CC SN events, 2312 of Type IIP and 916 of Type Ibc.
The asymmetry in training samples for the two classes causes our predictions for SNe Ibc to be worse than for SNe IIP (see Sections \ref{s:results}).

The PLAsTiCC dataset is not ideal for the eventual goal of strategizing real-time follow-up. 
As mentioned above, a major reason for this is the imbalance in number of LCs for each transient type. 
Additionally, the accuracy of some of the LCs is a subject of debate (M. Modjaz, personal communication) and needs to be explored further. 
PLAsTiCC serves as a convenient starting point for demonstrating essential components of the system and identifying limitations. We leave improving our reference library as an avenue of future development (see also discussion in Section \ref{s:conclusions}).

We use the reference library for training. In this paper we use it to set the value of the hyperparameter `k', which is number of nearest neighbours to use for predicting the future behavior of an event.
Results for testing are shown in Section \ref{s:results}.

\subsubsection{Convolutional Auto-Encoder}

\begin{figure*}
\begin{center}
\includegraphics[width=\textwidth]{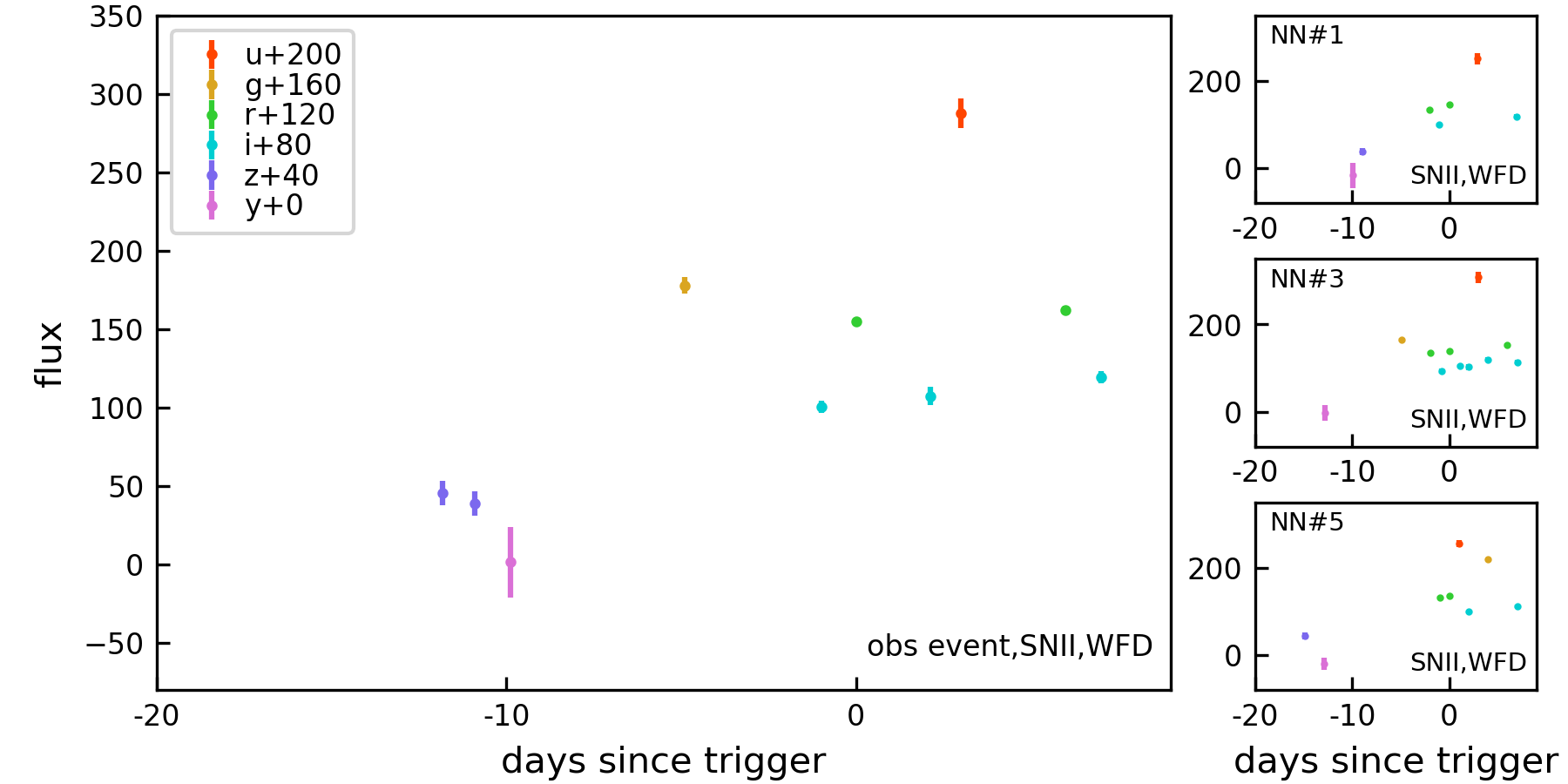}
\end{center}
\vspace{-8pt}
\caption{
Example of reference LCs matched to a given live event. The main panel shows a SN II event from LSST's WFD survey at 9 days since trigger. Panels to the left show three similar reference events (nearest neighbours 1, 3, and 5 from top to bottom) identified.
To select similar events we use the observed LC, event photometric redshift, and classification from a weak hypothetical classifier.
\label{f:kNN}}
\end{figure*}

Events within the window of action are passed to a CAE to find vector representations for them.
CAEs are a subset of dimensionality reduction methods that use convolutional neural networks (CNNs) to find vector representations of input data summarizing relational information. 
For example, given an RGB image, layers in a well-trained CNN find edges or color blobs.
CNNs are equivariant to translation, \citep[but not scale and rotation,][]{Goodfellow-et-al-2016}, i.e. they are agnostic to the location of an object in an image. 
This is especially useful because we would like to characterize an event without knowledge of its evolutionary phase (position on time axis) and true brightness (position on flux axis).

We adopt ResNet50 \citep{2015arXiv151203385H} minus its fully connected (FC) top layer trained on ImageNet \citep{ILSVRC15} as our CAE. 
Recent investigations have shown that features learned by layers in trained deep neural networks generalize well to even distantly related tasks
\cite[][]{2014arXiv1411.1792Y}.
Ideally, a dedicated CAE would be designed, as our current CAE is sensitive to the number and spacing of data points in a LC.
Training pre-trained ResNet50 plus a de-convolution block with LC pairs having the same shape but different photometric sampling should desensitize the CAE to survey cadence. 
However, for the purpose of this paper our current CAE is sufficient.

In order to supply LCs to our CAE, we first convert each PLAsTiCC event time-series in 6 LSST channels to 20 3x200x200 vectors. 
For this we bin the LC in each LSST channel in 200 bins in both time and flux. We bin flux by evaluating the normal distribution for flux mean (with value 1) and uncertainties at bin centers.
The flux range is fixed from -100 to 500 flux units. 
This choice spans the flux range for $98\%$ of CC SNe in the PLAsTiCC dataset.
The range in time is fixed from 20 days before trigger to the nearest integer day of the time since trigger for that event.
We then pass the 200x200 vectors in 3 LSST channel combinations to our CAE. 
The outputs from our CAE are stacked (always in the same order) and passed to system's matching component which is described next.

\subsubsection{Light-Curve Matching}

\rft\ uses CAE representations of ongoing events to find similar events in its reference library. 
It always computes L1 distances between CAE representations of live events and representations of library events at the same epoch. 
If photometric redshift is available from brokers, it additionally computes L1 distances between scalar redshift values for live and library events. 
Therefore, given a live event, similarity scores for every library event are computed using the mean z-score of CAE representation and redshift distances.
We find that L1 performs better L2 as a similarity metric. 
This is expected since L1 is known to perform better than L2 in high dimensions \citep{aggarwal01}. We leave an exploration of the use of more sophisticated similarity metrics (e.g. PSNR, FSIM, SSIM) to future work.

\rft\ also incorporates classification for events from brokers by performing a Bayesian update to its belief of the most similar library events for that event.
Specifically, given classification $c_\kappa$ for an event from classifier $\kappa$, it performs an update to its prior belief about the probability that the event actually belongs to class $C$ to:

\begin{equation} \label{e:update}
P(C|c_\kappa) = \frac{P(c_\kappa|C)P(C)}{P(c_\kappa)}
\end{equation} 
where
\begin{equation} 
\begin{split}
P(c_\kappa) = P(c_\kappa|C)P(C) + & P(c_\kappa|\bar{C})P(\bar{C}) \\ 
& + P(c_\kappa|\overline{C,\bar{C}})P(\overline{C,\bar{C}})
\end{split}
\end{equation} 

Here $P(C)$ is the \rft's prior probability that an event belongs to class $C$, $P(\bar{C})$ that it belongs to class $\bar{C}$, and $P(\overline{C,\bar{C}})$ that it belongs neither to class $C$ nor class $\bar{C}$.
Similarly, $P(c_\kappa|C)$ is the likelihood that an event actually belongs to class $C$, given this classification by classifier $\kappa$, and so on. 

For a given event, we define $P(C)$ as the fraction of events of class $C$ among `k' most similar library events. $P(\bar{C})$ is $1-P(C)$ and $P(\overline{C,\bar{C}})$ is 0. 
For this paper, this means that we always assign an event to either one of two CC SN types.
We note that such an assumption is very common in classifiers, including for most submitted in response to PLAsTiCC, i.e. they always assign an event to labels they train for.

In this paper, we assume classification information at every epoch is available via brokers from a hypothetical classifier ($\kappa$) with the following performance \citep[metrics motivated by][]{2019AJ....158..257B}:
$P(II_\kappa|II) = 0.6$, $P(II_\kappa|Ibc) = 0.1$, $P(Ibc_\kappa|Ibc) = 0.5$, and $P(Ibc_\kappa|II) = 0.05$.
We note that currently these metrics represent the most optimistic scenario since they correspond to a classifier looking at complete LC information in many cases.
In addition, in a real classifier these metrics would change as a function of time. 

Figure \ref{f:kNN} shows an example of similar library events matched to a live SN II event from LSST's WFD survey at 9 days since trigger. 
The system was also provided the event redshift and class (from classifier $\kappa$).
The system successfully exploits even sparse LC information to select similar library events.
The live event is a library event and chosen to demonstrate \rft's training in this and the next Section. 

\begin{figure}
\includegraphics[width=0.95\columnwidth]{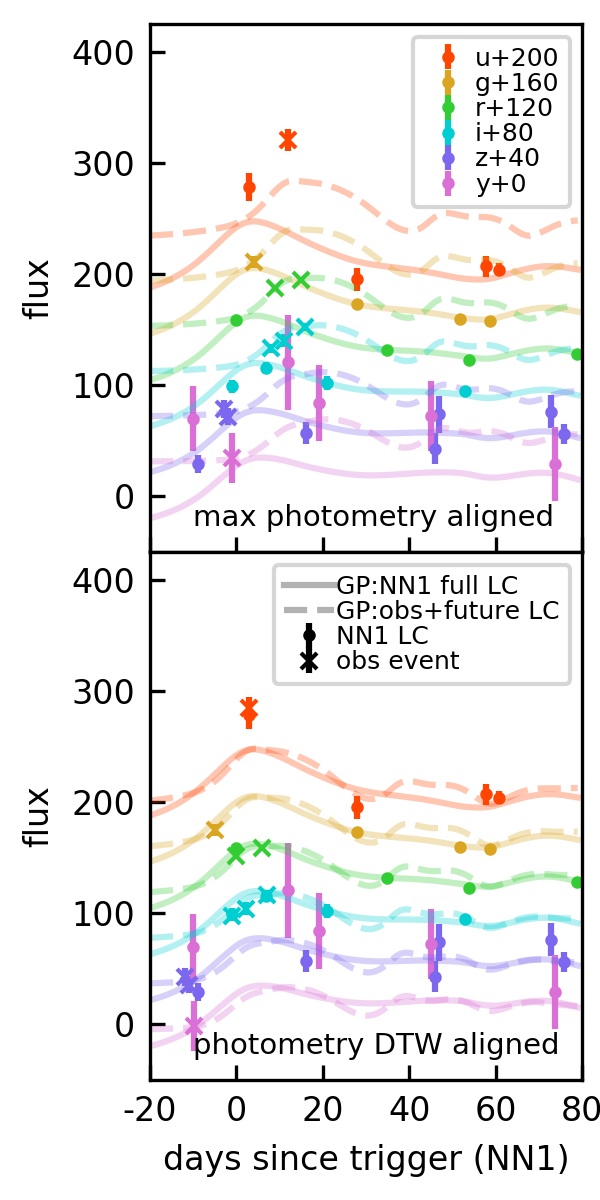}
\vspace{-5pt}
\caption{
Comparison of the effect of using max photometry anchoring (top) and Dynamic Time Warping (DTW, bottom) to align live observed and reference LCs. The observed LC (crosses) is for the same event as in the main panel of Figure \ref{f:kNN}. Dots show the full LC of the most similar reference event (upper-right panel of Figure \ref{f:kNN}).
Note that the underlying fits are only shown to guide the eye and not used for alignment.
\label{f:DTW}}
\end{figure}

\begin{figure*}
\begin{center}
\includegraphics[width=\textwidth]{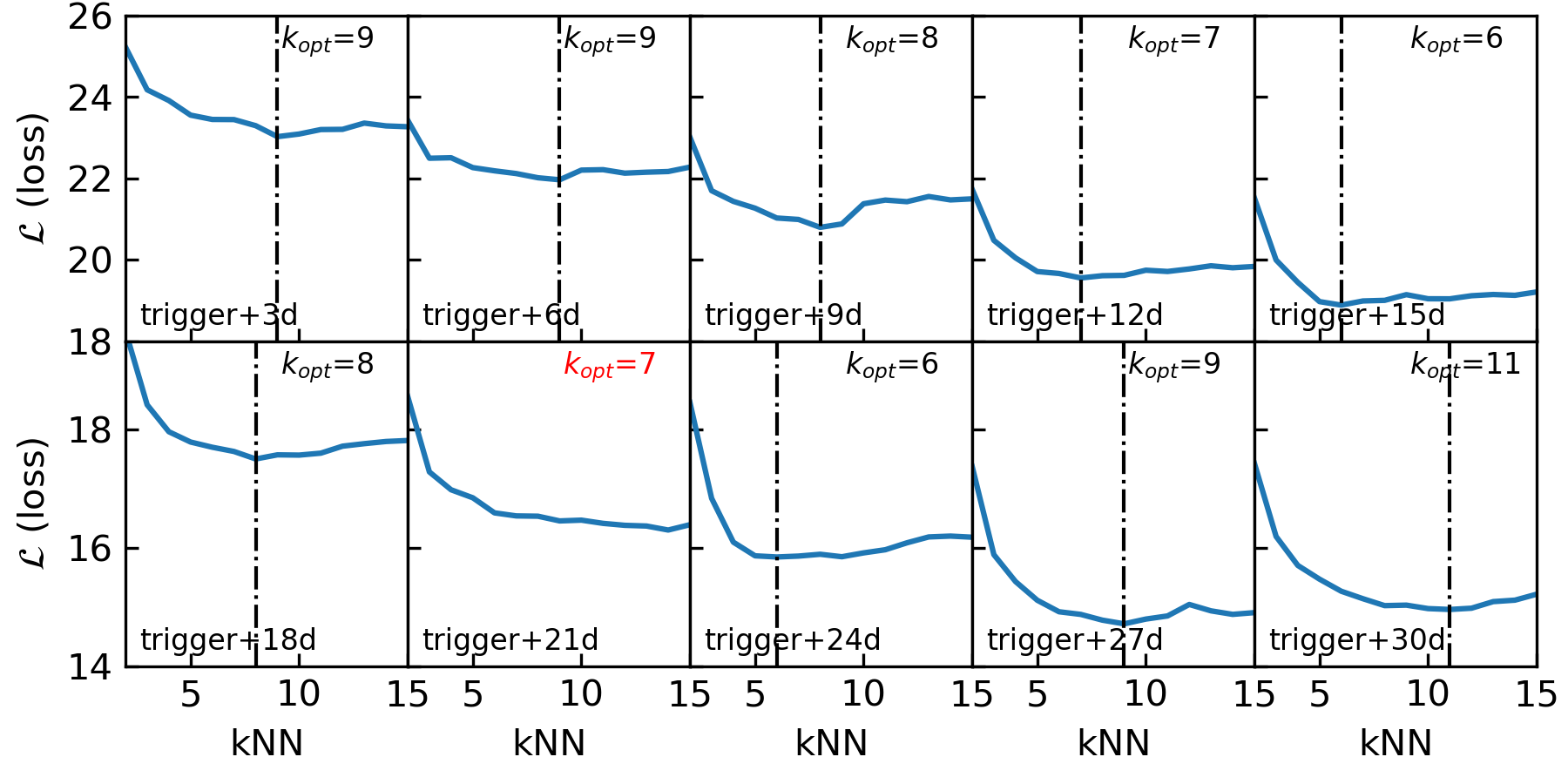}
\end{center}
\vspace{-15pt}
\caption{
Leave-one-out cross-validation loss as a function of `k' nearest neighbours used for prediction at various integer epochs since LSST trigger. 
Optimal values of `k' ($k_{opt}$) that correspond to minimum loss at every epoch are indicated.
Our loss at 21 days since trigger did not converge so we hard set the $k_{opt}$ at this epoch to 7 assuming physical continuity from before and after.
\label{f:LOOCV}}
\end{figure*}

\subsubsection{Light-Curve Prediction: Alignment and Regression} \label{sss:alignregress}

We use similar events in the reference library to predict the future LC for every live event within the window of action.
There are two steps to do this: alignment and regression.

First, the observed and similar reference events are aligned using Dynamic Time Warping (DTW).
DTW finds the optimal pairing between pairs of points in two time series, such that their Euclidian distance is shortest. 
DTW alignment is robust to signal stretching. 
This is particularly useful for aligning LCs that could differ slightly (e.g. due to different progenitor or explosion properties).
We use the \code{metrics} module of \code{tslearn} package \citep{tslearn} to perform DTW alignment. 
We exclude noise-level photometry when aligning by excluding flux measurements with mean flux values $<10\%$ of maximum observed flux in both live and reference LCs. 
We then use the median difference in time and flux between pairs of DTW aligned photometry to align the two LCs. 

Figure \ref{f:DTW} shows the result of using DTW to align sparse LSST LCs.
As expected, max photometry alignment can be very poor for sparse signals. DTW achieves good alignment without first needing to perform regression for re-sampling as would be needed to perform alignment using cross-correlation. 
We compare the performance of DTW to cross-correlation alignment in Appendix \ref{a:DTWvsCC}.

Next, we perform Gaussian Process (GP) regression for reference LCs following the method introduced by \citet{2019AJ....158..257B}. Briefly, we assume that LCs are described by a Gaussian process with mean 0 and a 2-dimensional Matern-3/2 kernel. As in \citet{2019AJ....158..257B} we fix the length scale in wavelength to 6000\AA. The length scale in time is estimated via maximum likelihood.
In this work, when flux estimated by GP is $<0$ we set its value to 0. We also assume a saturating flux at $1000$ units. The fits shown in Figure \ref{f:DTW} are GP fits.

We use both steps, alignment and regression, for each reference LC to estimate the future LC for an observed event. 
Specifically, we estimate the mean and standard deviation of predictions from `k' most similar reference events. 
We ignore uncertainties from GP fits as these are quite small. 

\newpage
\subsection{Training} \label{ss:train}

\begin{figure*}
\begin{center}
\includegraphics[width=\textwidth]{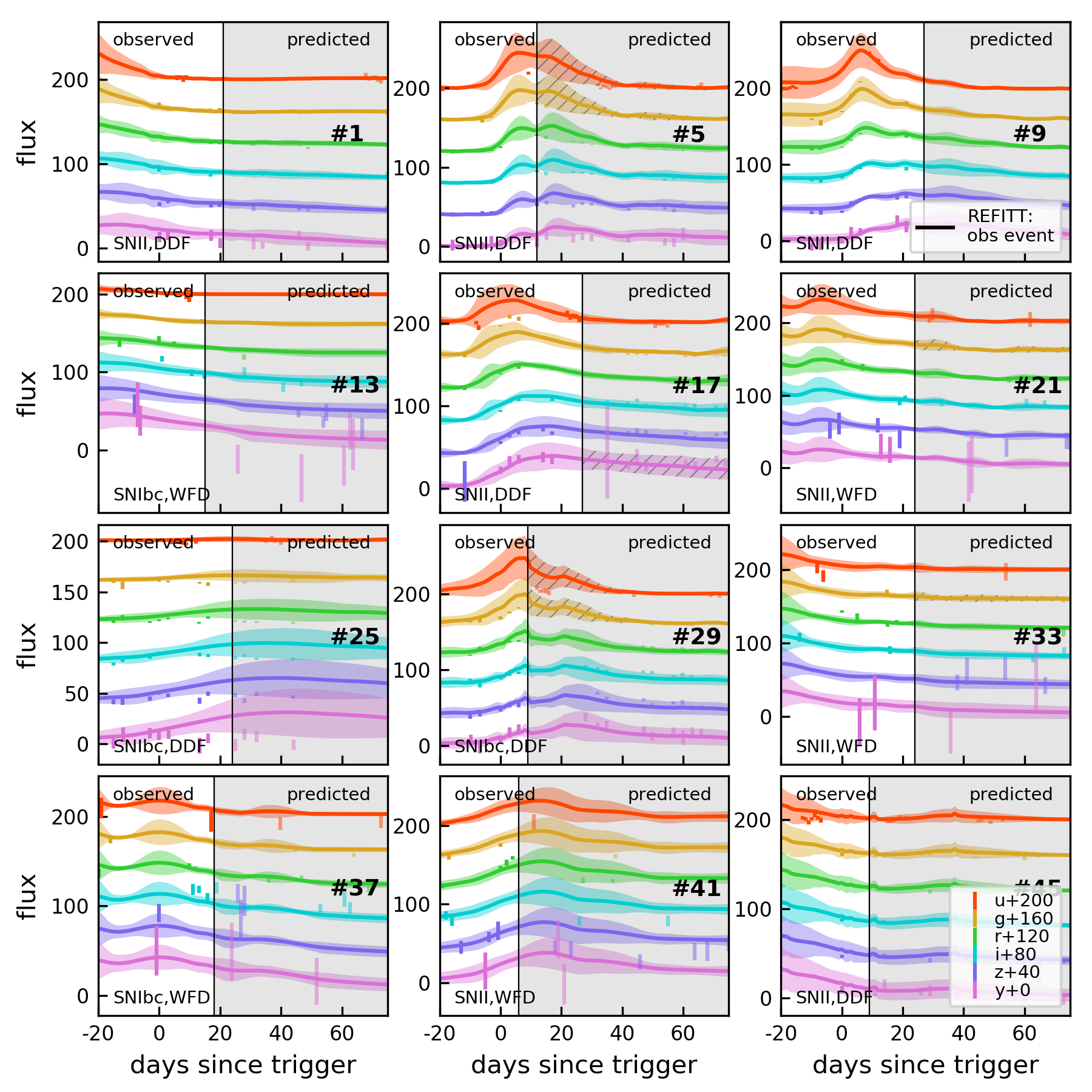}
\end{center}
\vspace{-15pt}
\caption{
Simulation of \rft's prediction, prioritization, and follow-up recommendation (hatched regions) on a random night of LSST operations.
\rft\ received $\sim 32000$ CC SN events within the pre-defined window of action, and ranked them based on its prediction quality (rank in bold; SN sub-type and survey type in bottom left corners).
For a given event, photometry in white regions are observed and grey regions are yet to be observed. Solid lines and shaded regions are mean and 1-$\sigma$ uncertainties of predictions.
Prediction uncertainties typically span the true error in predictions.
\rft\ does not require good LC coverage in order to make good predictions. 
The system makes good predictions for the event ranked 41 more than 6 days before peak in the u-band and for the event ranked 37 even though its peak is missed entirely.
\label{f:refitt_day}}
\end{figure*}

Our architecture requires the setting of the hyperparameter `k', where `k' is the number of most similar reference events to use for predictions. 
We estimate `k' that minimizes the training loss:
\setlength{\abovedisplayskip}{0pt}       
\begin{equation} \label{e:loss}
\mathcal{L}(k,t)=\frac{1}{n(\mathcal{D})}\sum\limits^{\forall\mathcal{D}}
\frac{\sum\limits_{i=1}^{N_{\rm future}}|\sigma_{i,pred}-|\mu_{i,pred}-\mu_{i,true}||}{N_{\rm future}}
\end{equation}
We compute a dedicated optimal `k' for every integer day since trigger.
Here $\mathcal{D}$ denotes the reference library. 
The effect of the first summation is performing leave-one-out cross-validation (LOOCV).
In the second term, for a given library event observed at a given time since trigger ($t$), we compute the mean absolute difference between the standard deviation of the signal prediction ($\sigma_{pred}$) from the absolute error of signal prediction at every future epoch with available photometry ($N_{\rm future}$).
The absolute error of the prediction is the absolute difference between the mean of the predicted flux ($\mu_{pred}$) and the mean of the true observed flux ($\mu_{true}$). 
In other words, $\mathcal{L}$ has the effect of selecting `k' such that, on average, \rft's predicted signal uncertainty is close to the true error in its prediction.
We only compute loss for events where $\mu_{pred}$ changes by at least 25 units in at least three LSST bands. This is to exclude low flux events that tend to have small $\sigma_{pred}$ values and flat loss curves.

Figure \ref{f:LOOCV} shows the LOOCV loss as a function of `k' most similar reference events used for prediction at various integer values of time since trigger. As expected the overall loss decreases with increasing time since trigger as more photometry become available and predictions improve.
At 21 days since trigger, our model is underfit, i.e. increasing `k' decreases bias without increasing variance. 
We hard set the optimal `k' at this epoch to 7, assuming physical continuity from epochs before and after.

To ensure robustness of the system to finding similar library events for prediction given noise, we carried out tests where we artificially increased or decreased flux uncertainties in LCs.
The system is robust to this source of noise: for changes within $\sim$25\%, the system selects the same neighbours for prediction. For changes up to $\sim$50\%, at most one neighbour differs.

Finally, we use the optimal `k' values to compute the mode of predicted LC uncertainties expected for the next day (using 1 flux unit bins) in each band. The modal uncertainty is the typical uncertainty due to intrinsic phase space scatter in the library. We use this metric to strategize follow-up (discussed next).

\subsection{Recommendations} \label{s:recommendation}

\rft\ makes a prediction for every event within its window of action. 
It then uses these predictions to design an observing strategy aimed at optimally augmenting the LCs.
In this paper, we use a data-driven metric for strategizing follow-up (see also Section \ref{s:conclusions}).

Each LC prediction is assigned a quality score based on how well it is known to have performed and expected to perform:
\begin{equation} \label{e:score}
\mathcal{S}=\left(\sum_{i}^{\rm N_{obs}}\frac{|\mu_{i,pred}-\mu_{i,obs}|}{N_{obs}}\right)+\alpha \left(\sum_{j=-20}^{80}\sigma_{j,pred}\right)
\end{equation}
where indices $j$ denote integer days since trigger and $\alpha$ is a weighting factor.
This score rewards predictions that have small mean observed errors and small overall uncertainties (from 20 days before up to 80 days after trigger).
In this work we set $\alpha$ to 1, however, it can be tuned to prioritize either metric.
We only score events whose mean predicted flux changes by at least 25 flux units in at least one band between 20 before and 80 days after trigger. This criterion is used to exclude events in their late fading phases and is arbitrary. 
We also only score events with $>0.25$ mean number of photometry per day to exclude events with several missing observations.

\begin{figure}
\includegraphics[width=\columnwidth]{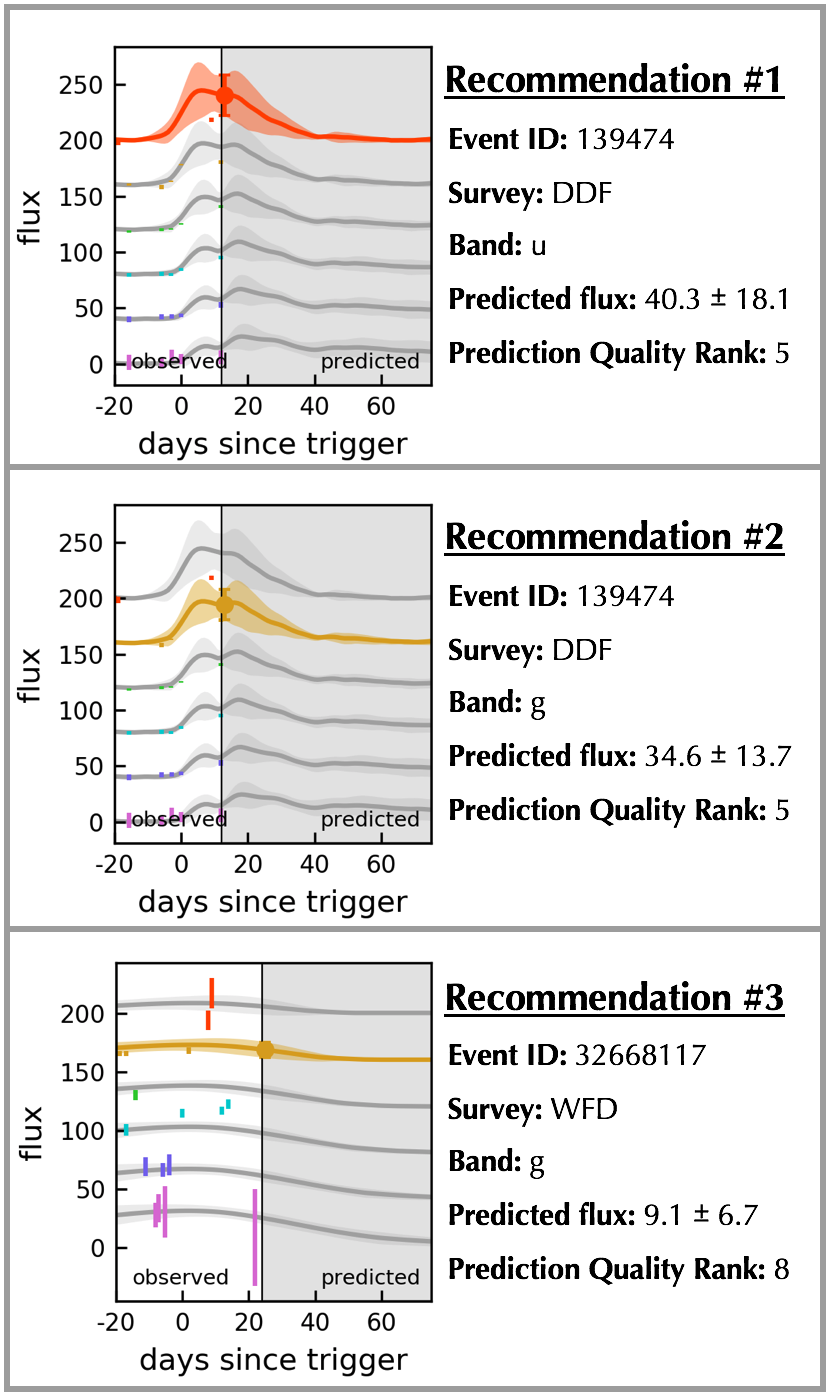}
\caption{
Example of \rft's follow-up recommendations for the same LSST night as in Figure \ref{f:refitt_day}. For this figure we do not show photometry yet to be observed.
Follow-up is recommended when the prediction uncertainty is higher than typical prediction uncertainties expected from training.
Note that the top two recommendations are for the same event, with priority for the band with larger prediction uncertainty.
\label{f:recommend}}
\end{figure}

\begin{figure*}
\begin{center}
\includegraphics[width=\textwidth]{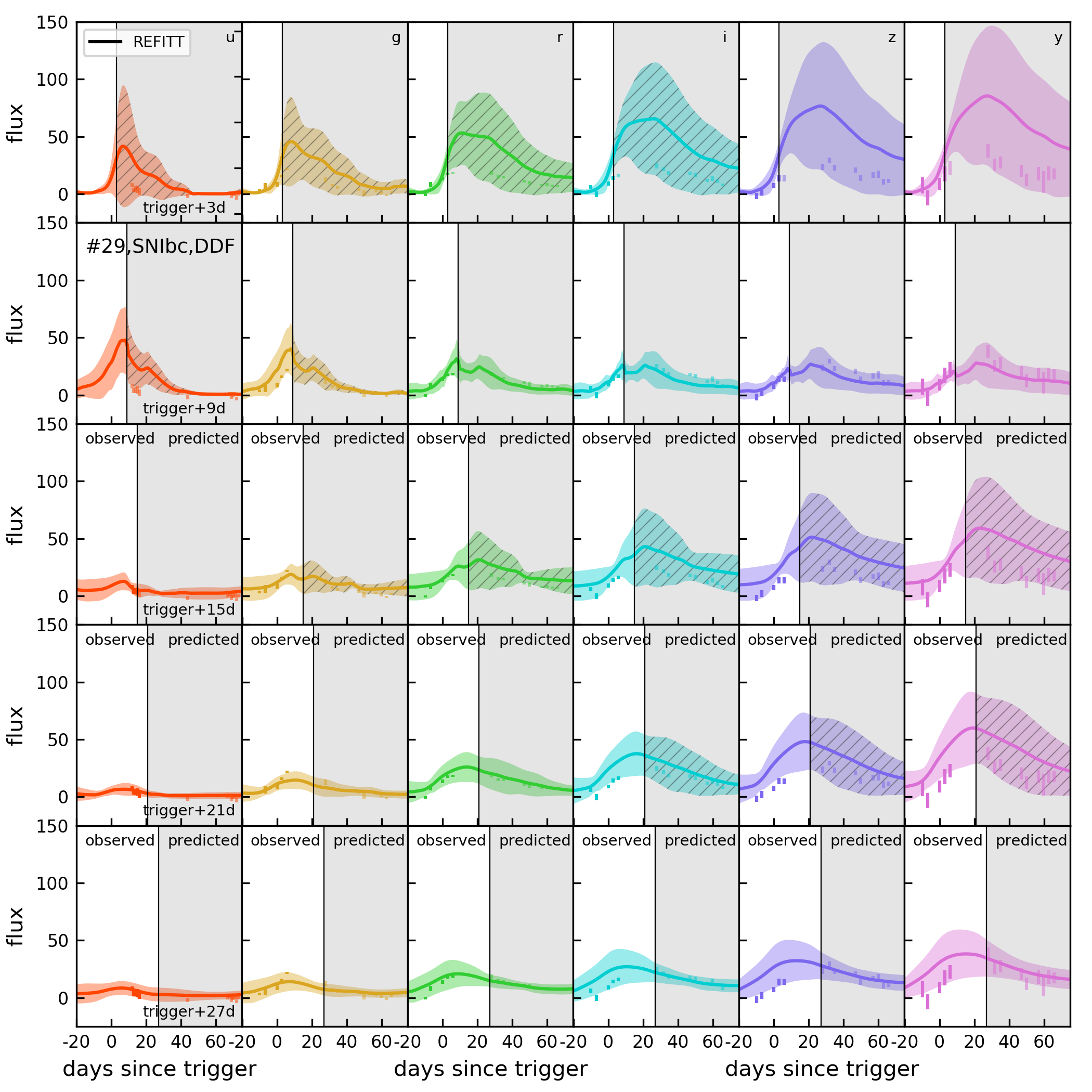}
\end{center}
\vspace{-10pt}
\caption{
\rft's prediction for a SN Ibc event from the DDF survey as a function of time since trigger (increasing downwards). 
As in Figure \ref{f:refitt_day}, photometry in white regions are observed and grey regions are yet to be observed. Solid lines, shaded and hatched regions have the same meaning as in Figure \ref{f:refitt_day}.
Panels from left to right separate observations and predictions by band (u to y).
This event is ranked 29 based on the prediction quality when at 9 days since trigger in the simulation shown in Figures \ref{f:refitt_day} and \ref{f:recommend}.
\rft's predictions improve and its uncertainty decreases with time as new photometry is collected.
\label{f:refitt_improvement}}
\end{figure*}

All events in the window of action are ranked according to $\mathcal{S}$.
We then search down the ranked list to check whether the predicted uncertainty for the next day in any band is greater than modal uncertainty computed during training. Since the modal uncertainty represents the bulk of the scatter in predictions due to the library, uncertainties larger than it represent variations between neighbours themselves. 
After confirming that an observation at that epoch and band will not be available given LSST's and other survey schedules (in this paper whether data is available at that epoch in the event simulation), we flag it for follow-up in order of $\mathcal{S}$. If more than one band in an event meets the above criteria we prioritize the band with the larger prediction uncertainty.

\section{\rft\ in the era of LSST} \label{s:results}

\begin{figure}
\includegraphics[width=0.97\columnwidth]{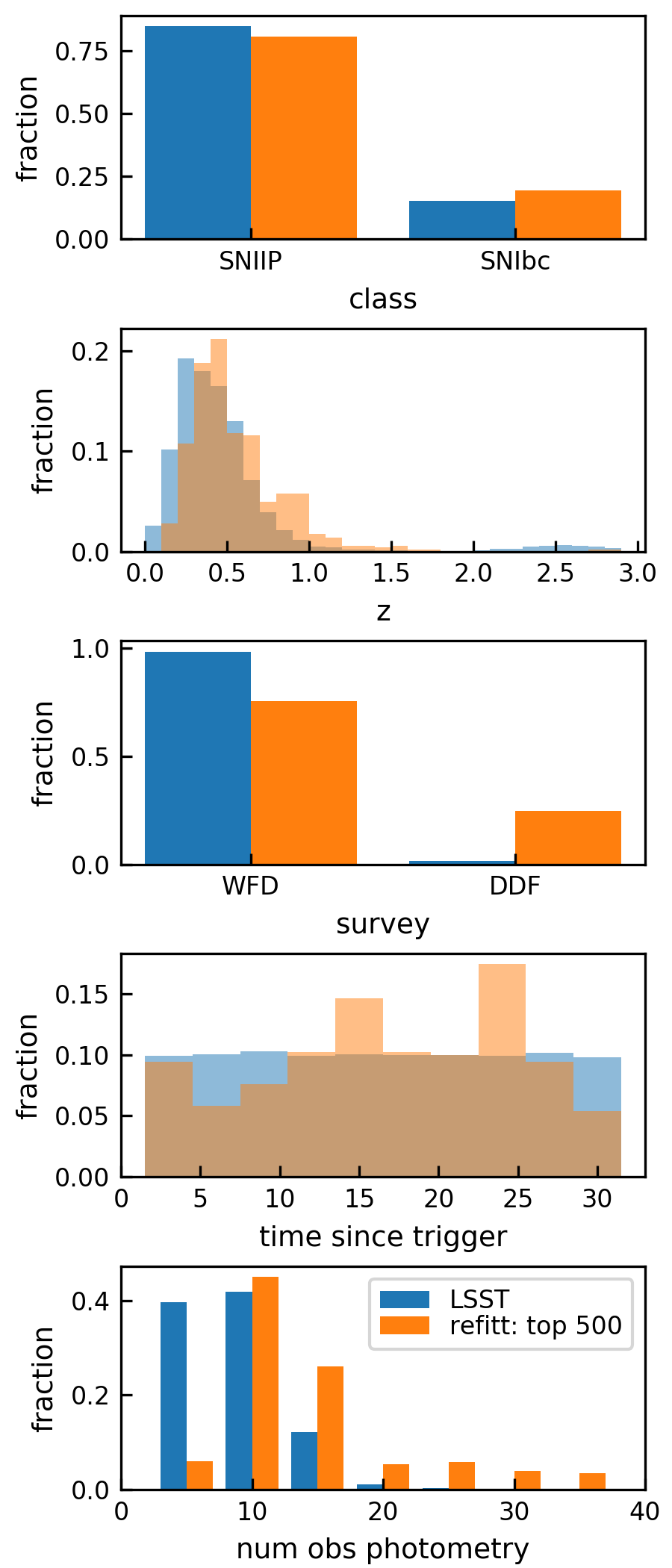}
\vspace{-2pt}
\caption{
Distribution of properties of top 500 events according to \rft's prediction quality (orange) compared to those from LSST (blue) on the same night as in Figures \ref{f:refitt_day} through \ref{f:refitt_improvement}.
\rft\ does not exhibit a strong bias in the types of events it prioritizes for follow-up, however exhibits mild preference for events from the DDF survey and correspondingly more photometric points in an event. This is mostly due to the over-representation of DDF events in its reference library (due to our selection criteria listed in \ref{ss:lib}).
\label{f:refitt_dist}}
\end{figure}

Figures \ref{f:refitt_day} and \ref{f:recommend} demonstrate \rft's predictions, prioritization, and, follow-up recommendations on a random night of LSST operations.
We simulate the alert stream using CC SN events in the PLAsTiCC dataset not in the reference library (i.e. test set), assuming they occur randomly over the 3 year LSST operation simulated in PLAsTiCC. 
We simulate broker input by supplying event photometric redshift and class (from classifier $\kappa$).
On a given night, the system considers $\sim32000$ CC SN events that are within the 30 day window of action.

\rft\ is able to make good predictions for both SN II/Ibc subtypes and events from WFD/DDF surveys.
The 1-$\sigma$ uncertainties typically span the true error in predictions which indicates successful training. 
\rft\ does not require good LC coverage in order to make good predictions. 
For example, the system makes good predictions for the event ranked 41 in Figure \ref{f:refitt_day} more than 6 days before peak in the u-band. Similarly, the system makes good predictions for the event ranked 37 even though its peak is missed entirely.
For the event ranked 25 in Figure \ref{f:refitt_day}, an observation is not recommended despite the uncertainty in several bands being greater than the system's tolerance (modal uncertainty from training) because LSST is expected to revisit the object within the next 24 hours.

Generally, the system's predictions for SNe II are better than Ibc. 
This is due to the imbalance in the number of SN II and Ibc events in the reference library.
Our reference library has more SN II than SN Ibc events by a factor of $\sim$2.5 (see Section \ref{ss:lib}).
Developing a reference library with uniform distribution for event properties would be crucial for the system to operate live and will be a central focus of future work (see Section \ref{ss:lib} and Section \ref{s:conclusions}).


Figure \ref{f:refitt_improvement} demonstrates how \rft\ updates its predictions and follow-up recommendations given new data.
As expected, \rft's predictions improve over time. The effect of new data can be clearly seen when the event shown is at 15 days since trigger, when the first u-band photometry become available.
We note that the uncertainty at 3 days since trigger for this event is not typical for all events and can be better, especially for SN II events.
Similarly, the fit achieved by 27 days since trigger is not typical and can be better for SN II (see examples in Figure \ref{f:refitt_day}) and worse for SN Ibc events. 
As discussed earlier, these differences are due to the imbalance between the number of reference events of the two SN subtypes in the reference library. 

\rft\ also updates its follow-up strategy given new data.
In the early phases of the event shown in Figure \ref{f:refitt_improvement}, the system characterizes it as a SN II (bluer photometry) and recommends follow-up in u, g, r, and i bands. 
Then given new data in g, r, and i bands over the next 6 days \rft\ is able to improve its prediction and no longer recommends follow-up in r and i bands, though still characterizing the event as SN II.
After 6 more days, given the first u-band data, the system begins characterizing the event more like a SN Ibc (redder photometry) with an increased uncertainty in its predictions. It also recommends follow-up in all bands from g through i.
We underscore that the ideal confirmation of \rft's predictive or learning ability would be with live data.
Machine learning models learn to represent properties of training data. 
If these differ from real data in ways such as to be critical for the task the model is performing, it would respond poorly to real data despite performing remarkably well during training.

Figure \ref{f:refitt_dist} shows the distribution of properties of the top 500 events according to \rft's prediction quality compared to those from the LSST survey.
\rft\ does not have a strong bias for either SN II/Ibc events or WFD/DDF surveys,
The latter is desirable because it means that system does not require high cadence photometry in order to make good predictions. 
Focusing on DDF events for augmentation might not be the most productive use of follow-up resources as they may have enough photometry to allow them to be interpreted with theoretical models.
The system also does not have a preference for late phase events (or longer times since trigger).
This naturally favors data collection at earlier more profitable phases of an event.

\section{Conclusions} \label{s:conclusions}

LSST's unprecedented high quality but sparsely cadenced photometry requires real-time follow-up to maximize the science potential of the survey. Given the high cost for obtaining data, follow-up strategies need to be value-driven as not all data are equally valuable for constraining science. 
We call systems that autonomously strategize and co-ordinate follow-up in real-time ORACLEs. 
\rft\ is a prototype of an ORACLE that uses predictive modeling for sparse multi-channel LSST LCs and value-based metrics to strategize follow-up in real-time.
The system capitalizes on, without interfering with, survey schedules and is flexible to accommodate any science objective.
It also seeks to optimally leverage global networks of telescopes (e.g. AEON, LCO \citep{2013PASP..125.1031B}, GROWTH \citep{2019PASP..131c8003K}) for optimally augmenting valuable data.

The current implementation of \rft\ can predict sparse LSST CC SN LCs and use them to design follow-up strategies. 
However, in order to operate on the full LSST survey stream multiple improvements are desirable.
First is the development of a robust reference library with LCs for diverse transient types, 
along with multi-wavelength photometric and spectroscopic templates \citep[e.g.][]{2019MNRAS.489.5802V}. 
Second, is development of a dedicated CAE, potentially by fine-tuning the ResNet50 architecture \citep{DBLP:journals/corr/ZeilerF13,Oquab:2014:LTM:2679600.2680210,2014arXiv1411.1792Y}.
Third, the system should ingest additional information from brokers (e.g. Milky Way extinction, galaxy association) to improve its predictions since it is more likely that brokers will reliably be able to provide contextual information compared to real-time classification. These can be included as additional parameters when scoring reference event matches.

Finally, it may be beneficial to strategize follow-up to optimally constrain theoretical models of interest. 
One option would be to prioritize follow-up in order of increasing D-optimality \citep{kiefer1959} of obtaining an additional observation at the next available epoch. 
Such an approach would maximize the Shannon information \citep{6773024} about parameters describing a chosen theoretical model for the full alert stream.



\acknowledgments
We are grateful to our anonymous referee for their careful and insightful review.
We thank Tom Matheson for providing useful inputs that helped improve the manuscript.
NS thanks Michael Katz and John Banovetz for helpful discussions. 
NS acknowledges the IDEAS program (\url{ideas.ciera.northwestern.edu}) for her training in statistics and machine learning. This research was supported in part through computational resources provided by Information Technology at Purdue, West Lafayette, Indiana.

\software{\code{tslearn} \citep{tslearn},
    \code{George} \citep{2015ITPAM..38..252A},
    \code{keras} \citep{chollet2015keras},
    \code{pandas} \citep{pandas},
    \code{scipy} \citep{scipy}    }

\newpage
\appendix

\section{Comparison of Cross-Correlation and DTW LC Alignment} \label{a:DTWvsCC}

A common technique used to align signals is cross-correlation. 
The method requires both signals to have measurements at all comparison timestamps. For irregularly sampled signals, this can be accomplished by estimating the signals at regular time intervals (resampling) using a fitting procedure. 

\begin{figure}
\includegraphics[width=0.95\columnwidth]{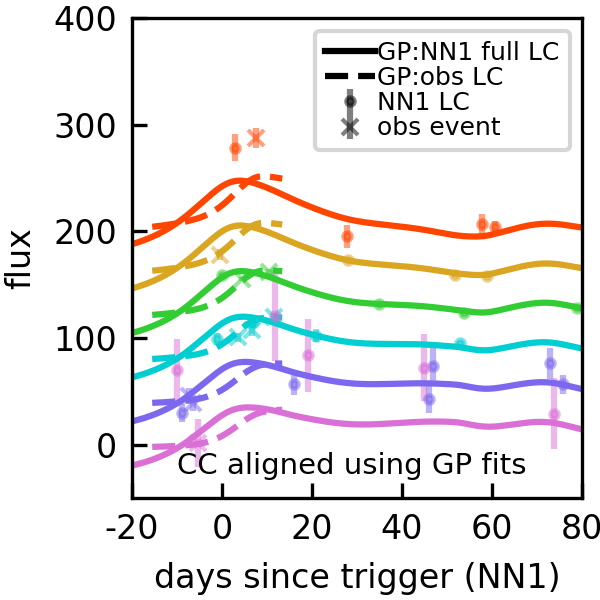}
\vspace{-5pt}
\caption{
Similar to Figure \ref{f:DTW} but using cross-correlation to align live observed and reference LCs. 
Photometry shapes (crosses and dots) have the same meaning as in Figure \ref{f:DTW}. 
Differently from Figure \ref{f:DTW}, the GP fits shown in opaque solid and dashed lines were used to resample the signals for alignment. 
Cross-correlation performs worse than DTW in achieving alignment.
\label{f:DTW_vs_cc}}
\end{figure}

Figure \ref{f:DTW_vs_cc} shows the result of using cross-correlation to align LCs. We use our GP regression to resample both LCs in 0.1 day intervals. 
Then the median optimal timestamp estimated from cross-correlating the resampled LCs in each band was used to align them along the time-axis.
Similarly, the median maximum flux from the resampled LCs in all bands was used to align them along the flux-axis.
The alignment obtained via this method is worse than from DTW (compare to the lower panel of Figure \ref{f:DTW}). It also takes 6-7 times longer to compute than DTW.
For partial LSST LCs, DTW yields superior performance, achieving faster and more accurate alignment without requiring LC resampling.

\bibliography{references-master}
\bibliographystyle{aasjournal}

\end{document}